# OCELOT: Direct Atmospheric Forecasting from Heterogeneous Earth Observations Using a Graph–Transformer Hybrid Model

Azadeh Gholoubi [a*], Ronald McLaren [b], Mu-Chieh Ko [b], Xin Jin [c], Nicholas Esposito [c], Andrew Collard [d], Cory Martin [d], Russ Treadon [d], Daniel Holdaway [d] and Daryl Kleist [d]

[a] *Axiom Consultants, in support of NOAA/National Weather Service/National Centers for Environmental Prediction/Environmental Modeling Center*

[b] *Lynker, in support of NOAA/National Weather Service/National Centers for Environmental Prediction/Environmental Modeling Center*

[c] *SAIC, in support of NOAA/National Weather Service/National Centers for Environmental Prediction/Environmental Modeling Center*

[d] *NOAA/National Weather Service/National Centers for Environmental Prediction/Environmental Modeling Center*

[*] *Corresponding author:* Azadeh Gholoubi, azadeh.gholoubi@noaa.gov

ABSTRACT: This study presents OCELOT (Observation-Centric Estimation and Learning for Outlook Trajectories), a global machine-learning forecasting system that predicts future Earth observations directly from heterogeneous satellite and in-situ measurements. Unlike data-driven weather models trained on gridded reanalysis states, OCELOT operates natively in observation space, preserving instrument-specific sampling, viewing geometry, and measurement characteristics. The system combines per-instrument graph-attention encoders, a shared spherical icosahedral latent mesh, a hybrid sliding-window Transformer/spatial graph neural network processor, and metadata-conditioned decoders to produce forecasts up to 12 h ahead. OCELOT is trained on observations for the years 2015 through 2023, validated on the year 2024, and evaluated out of sample on 2025 observations across satellite radiances, radiosondes, aircraft, and surface networks. In the 2025 evaluation, OCELOT produces spatially coherent +12 h forecasts across independent observing systems: microwave temperature-sounding channels show RMSE values of 1.24–1.87 K, while the more surface- and cloud-sensitive AVHRR infrared window channel shows a higher RMSE of 3.95 K. Vertical profile diagnostics show physically consistent radiosonde and aircraft temperature structure. Surface forecasts remain stable through 12 h, with 2-m air-temperature RMSE increasing from about 3.2 K at +3 h to about 3.6 K at +12 h. In paired observation-space comparisons, OCELOT remains less accurate than operational GFS but substantially outperforms persistence at longer lead times for 2-m temperature and 10-m wind components. These results demonstrate that observation-space forecasting can recover large-scale atmospheric structure and provide meaningful short-range skill without reanalysis supervision.

SIGNIFICANT STATEMENT: Most machine-learning weather prediction systems are trained on gridded analysis or reanalysis datasets, which are expensive to produce and inherit assumptions from numerical models and data-assimilation systems. Building on recent observation-space forecasting work, we present OCELOT, a modular NOAA research prototype that produces globally coherent short-range forecasts directly from heterogeneous Earth observations, including satellite radiances, scatterometer backscatter, radiosonde and aircraft profiles, and surface networks, without using reanalysis fields as training targets. OCELOT learns a shared latent representation on a spherical icosahedral mesh while preserving instrument-specific sampling geometry and metadata through graph-based encoders and decoders, providing a scalable pathway for reanalysis-independent AI weather prediction.

## 1. Introduction

Machine-learning-based global weather forecasting has advanced rapidly. Systems such as GraphCast (Lam et al. 2023), Pangu-Weather (Bi et al. 2023), FourCastNet (Pathak et al. 2022), Keisler (2022), and FengWu (Chen et al. 2023) demonstrate that neural networks trained on large-scale reanalyses can rival—and in some cases exceed—traditional numerical weather prediction (NWP) systems. These models learn mappings between gridded atmospheric states and their future evolution, yielding fast, skillful forecasts in deterministic and, more recently, probabilistic settings.

Despite these advances, such approaches depend fundamentally on gridded state representations produced by NWP or reanalysis systems. Multidecadal reanalyses are expensive to generate, limited in temporal coverage, and intrinsically tied to the assumptions, physics, and assimilation procedures of the underlying model (Kalnay 2003; Bannister 2017). Consequently, grid-based machine-learning forecasts inherit structural dependencies on assimilation cycles and state-reconstruction pipelines. This raises an important question: *can large-scale forecasting be performed directly from heterogeneous observations, without first converting measurements into gridded analysis states or learning an evolution operator for reanalysis-derived model states?*

Observation-space approaches learn the temporal evolution of raw measurements from satellite radiometers, geostationary imagers, radiosondes, surface networks, and aircraft reports. Forecasting in observation space avoids reliance on assimilation cycles, preserves instrument-specific sampling geometry and units, and enables direct evaluation of cross-instrument consistency and predictability, extending concepts developed in forecast sensitivity to observations and observation-impact studies (Langland and Baker 2004; Gelaro et al. 2010). GraphDOP (Alexe et al. 2024) was, to our knowledge, the first unified framework for global observation-space forecasting. However, the broader design space of scalable, modular, and extensible observation-space architectures remains largely unexplored.

Building on the observation-space forecasting paradigm demonstrated by GraphDOP, OCELOT (Observation-Centric Estimation and Learning for Outlook Trajectories) explores whether operational satellite and in-situ observing streams can support AI-based short-range forecasting directly in observation space, without first requiring gridded reanalysis or data-assimilation targets. Developed as a NOAA National Weather Service research prototype,

OCELOT combines instrument-specific observation encoding, a shared spherical latent mesh, temporal-spatial latent processing, and metadata-conditioned decoding to preserve native observing-system geometry while enabling cross-instrument information exchange.

This paper presents the first end-to-end description and out-of-sample evaluation of OCELOT v1. To provide operational context for the surface-variable results, we include diagnostic comparisons with the NCEP Global Forecast System (GFS), a mature operational physics-based NWP system initialized through cycled data assimilation, and with a same-location persistence baseline. Section 2 formalizes the observation-space forecasting problem; Section 3 describes the data and preprocessing pipeline; Section 4 details the architecture; Section 5 describes the training methodology and reproducibility settings; Section 6 reports 12-hour forecast results across instruments and diagnostic comparisons with GFS and persistence; and Sections 7–9 provide discussion, limitations and future directions, and conclusions.

### 1.1 Main Advances

The main advances of this paper are fourfold. First, we present a reanalysis-independent global forecasting framework that operates natively on heterogeneous Earth observations, with no gridded analysis used for model training or prediction. Second, we introduce the OCELOT v1 architecture: a modular encoder–processor–decoder system with per-instrument graph attention network v2 (GATv2) encoders and decoders, metadata-conditioned decoder queries, and a hybrid sliding-window Transformer/spatial-GNN processor on an icosahedral mesh. Third, we use a masked, instrument-rebalanced, channel-weighted mean squared error (MSE) objective designed for heterogeneous observing systems with large sample-count differences. Fourth, we provide an out-of-sample 2025 evaluation across seven instrument families, including native observation-space verification and diagnostic comparisons with GFS and persistence.

### 1.2 Related Work

OCELOT belongs to a small but growing family of end-to-end, observation-trained machine-learning weather models. The closest methodological precedent is GraphDOP (Alexe et al. 2024; Boucher et al. 2025; Lean et al. 2025), which demonstrated that global forecasts can be learned and initialized directly from heterogeneous Earth-system observations without gridded reanalysis inputs. GraphDOP further shows that observation-trained models can develop coherent latent representations of Earth-system structure and cross-instrument or cross-domain relationships.

OCELOT builds on this observation-space forecasting paradigm while exploring a different architectural and evaluation regime. Three distinctions from GraphDOP are central. First, OCELOT uses per-instrument bipartite GATv2 encoders and decoders on observation–mesh and mesh–observation graphs, preserving instrument-specific sampling geometry, footprint structure, and viewing metadata. Second, OCELOT uses decoder-query conditioning by projection, so that scan angle, pressure level, and valid-time metadata explicitly initialize target-node queries in the full hidden dimension before mesh-to-target decoding. Third, OCELOT evolves its latent state on a 40,962-node spherical icosahedral mesh using a hybrid sliding-window Transformer/spatial-GNN processor that interleaves temporal attention with explicit mesh-edge message passing. This differs from GraphDOP's latent processor, which uses sliding-window Transformer dynamics on an o96 reduced Gaussian latent grid, with graph mappings used primarily to encode observations into and decode predictions from the latent representation. Although GraphDOP is the closest methodological reference, we do not report a direct numerical benchmark against it in this first OCELOT v1 evaluation. Such a benchmark would require a common observation archive, matched observation types, identical forecast windows, shared target variables, and a consistent verification protocol. Published GraphDOP results are based on different training periods, observation sets, model capacity, latent-grid design, and medium-range evaluation targets, whereas OCELOT v1 is evaluated here as a short-range 0–12 h NOAA observation-space prototype. We therefore use GraphDOP as the primary conceptual and architectural reference and leave a controlled head-to-head comparison for future work.

Aardvark Weather (Allen et al. 2025) provides a complementary end-to-end approach that replaces the traditional NWP pipeline and produces gridded global and local-station forecasts, including a learned analysis stage. By contrast, OCELOT keeps the primary prediction target natively in observation space at the original sensor locations, valid times, and metadata coordinates. This allows verification directly against held-out measurements without first mapping the targets to a gridded analysis state, while still allowing the learned latent mesh to be queried for gridded diagnostic products.

## 2. Forecasting Objective and Notation

OCELOT addresses the task of short-range observation-space forecasting, in which future Earth observations are predicted directly from past observations without reliance on gridded model states or reanalyses. We index individual observations by $j$, so that the $j$-th observation is defined as

$$o_j = (x_j, t_j, m_j, y_j), \tag{1}$$

where $x_j$ denotes geographic location, represented by latitude and longitude; $t_j$ is the observation time; $m_j$ contains instrument-specific metadata, such as scan geometry, platform identifier, or pressure level; and $y_j$ is the measured quantity. Given the set of heterogeneous observations $O_{in}$ collected within the 12-h input window ending at time $t_0$, OCELOT predicts the value $\hat{y}_j$ for each target query ($x_j$, $t_j$, $m_j$), where $t_j \in (t_0, t_0 + 12\ h]$. During training and evaluation, the corresponding verifying observation $y_j$ is available for each target observation $o_j \in O_{tgt}$. We represent OCELOT as a parameterized forecasting function $f_\theta$, where θ denotes all trainable model parameters:

$$\hat{y}_j = f_\theta(O_{in}; x_j, t_j, m_j). \tag{2}$$

The prediction $\hat{y}_j$ is compared with the observed value $y_j$ during training and evaluation. Forecasts are decoded separately for each observing system and variable while sharing a common latent atmospheric representation. This formulation preserves the native spatial sampling, geometry, metadata, and measurement units of each instrument while allowing information to be exchanged across observing systems through the shared mesh.

Temporal evolution is modeled by latent-space rollout: the shared latent mesh state is advanced in $K$=4 steps of $\Delta t$ = 3 h each, with supervision at every step. Predicted observation values are used for loss computation but are not fed back into the encoder as new observations. Forecasting directly in observation space avoids interpolation and representativeness errors associated with gridding irregular observations, preserves instrument-specific geometry and metadata, and enables integration of observing systems with widely varying spatial density and sampling frequency. Although the latent mesh representation can be queried at arbitrary points to produce gridded-like diagnostic products, such products are not part of the learning objective; all training targets and loss functions are formulated strictly in observation space.

## 3. Data and Preprocessing

OCELOT ingests heterogeneous Earth observation datasets spanning a broad suite of satellite- and in-situ-based observing systems (Table 1).

| Observation type | Platform / sensor | Prognostic target variables | Predicted channels |
|---|---|---|---|
| Microwave sounders | ATMS, AMSU-A | Multi-channel brightness temperatures | 22, 15 |
| Passive microwave imagers | SSMIS | Brightness temperatures | 24 |
| VIS/IR imagers | AVHRR, SEVIRI | VIS/IR radiances and brightness temps | 3, 8, 8 |
| Scatterometer | ASCAT | Backscatter coefficient | 3 |
| Aircraft reports | AMDAR, AIREP | Air temp, zonal/meridional wind | 3 |
| Radiosondes | Global RAOB network | Temp, dew point, wind profiles | 4 |
| Surface observations | Land and marine | MSLP, T2m, TD2m, U10m, V10m | 5 |

**Table 1.** Observations used in OCELOT v1. For satellite instruments, "channels" refers to radiometric spectral or beam channels. For in-situ observations, the count refers to prognostic target variables or components. All listed variables, including surface mean sea-level pressure, are prognostic observation-space targets in OCELOT v1. Gridded mesh-query fields generated by querying the learned decoders on the latent mesh are diagnostic products only and are not used as training targets. Missing or invalid targets are masked and do not contribute to the loss. For satellite inputs, missing standardized feature values are filled with zero, and missing metadata entries are filled using available column means in transformed metadata space. For conventional inputs, missing feature values use fixed sentinel fills after clipping, and metadata are filled using available column means followed by deterministic physical scaling.

Note: ATMS, Advanced Technology Microwave Sounder; AMSU-A, Advanced Microwave Sounding Unit-A; SSMIS, Special Sensor Microwave Imager/Sounder; AVHRR, Advanced Very High Resolution Radiometer; SEVIRI, Spinning Enhanced Visible and Infrared Imager; ASR/CSR, all-sky/clear-sky radiance; ASCAT, Advanced Scatterometer; AMDAR, Aircraft Meteorological Data Relay; AIREP, Aircraft Reports; PIREP, Pilot Reports; MDCRS ACARS, Meteorological Data Collection and Reporting System Aircraft Communications Addressing and Reporting System; TAMDAR, Tropospheric Airborne Meteorological Data Reporting; RAOB, radiosonde observation; MSLP, mean sea level pressure; T2m, 2-m air temperature; TD2m, 2-m dew point temperature; U10m/V10m, 10-m zonal/meridional wind components; VIS/IR, visible/infrared.

All datasets are accessed in the cloud-optimized Zarr format, enabling efficient sequential or randomized access to multiyear global archives without loading the full dataset into memory. The Zarr archives are generated from operational NOAA Global Data Assimilation System (GDAS) observation dumps, including satellite radiance, surface, radiosonde, aircraft, and scatterometer data streams. GDAS is used here only as a source of quality-controlled observational records and metadata; OCELOT does not ingest GDAS analyses, reanalysis fields, model states, or assimilation increments during training or inference.

Observations are stored chronologically in cloud-optimized Zarr archives and are accessed through logical 12-hour time-window indexing at runtime. Each sampled training example contains a 12-hour input observation set and the subsequent 12-hour target interval used for multi-step forecasting via latent rollout. The target interval is subdivided into four 3-hour target windows corresponding to the +3, +6, +9, and +12 h forecasts; target observations are not duplicated into separate stored files. Per-instrument input features include normalized measurement values, instrument-specific metadata, sensor- and solar-geometry angles where applicable, geolocation encodings based on sine/cosine transformations of latitude and longitude, and temporal encodings representing time of day and seasonality.

### 3.1 Observation Archive and Time Splits

We partition eleven consecutive years of global observations into separate training, validation, and held-out test periods. For each observation type, the data are stored in a single cloud-optimized Zarr archive, and the split is applied logically through time-based indexing rather than by physically separating files. OCELOT is trained on 2015-2023 observations, validated on 2024 observations for early stopping, learning-rate scheduling, and model selection, and evaluated out of sample on 2025 observations. The 2025 test period is fully held out and is not used during training, validation, normalization, or hyperparameter tuning.

To scale to decade-long datasets without excessive memory use, OCELOT reads observations at runtime using windowed temporal indexing; full-year or multi-year tensors are never loaded into memory. During training, a random-window sampling callback selects a new 12-day window at each epoch to improve seasonal coverage, while validation uses a fixed sequential 12-day window that is advanced every N epochs.

### 3.2 Quality Control, Subsampling, and Feature Normalization

Quality control is applied across satellite and conventional observing systems using observation-level checks defined in the OCELOT configuration. For satellite instruments, OCELOT uses instrument-provided quality flags and channel-specific validity checks where available. For conventional observations, including surface reports, radiosondes, and aircraft reports, OCELOT applies provider-supplied quality flags together with physically reasonable range checks on pressure, temperature, dew point, wind speed, wind direction, and height or pressure-level metadata. Surface observations additionally use simple cross-variable consistency checks, including dew point not exceeding air temperature and limits on temperature-dew-point spread. These filters retain observations that pass the available QC indicators and physical plausibility checks, but they do not use variational assimilation, background-departure screening, analysis increments, or reanalysis-based filtering.

Following QC, OCELOT applies per-instrument, per-variable standardization using fixed normalization statistics held constant across training, validation, and evaluation. Angular metadata are transformed using cosine representations; geolocation and time are encoded using sine and cosine transformations. Invalid targets and inactive channels are retained in the data structure but excluded from learning and evaluation through explicit validity masks. During evaluation, predicted values are inverse-transformed to physical units.

Missing values are handled separately for feature channels and metadata. Observation values that fail quality control or are missing are marked invalid at the channel level; during training and evaluation, target-channel validity masks are propagated to the loss so that invalid target elements do not contribute to the objective or masked metrics. For model inputs, missing standardized satellite feature values are filled with zero, corresponding to the neutral value in standardized units, while conventional feature inputs use a fixed sentinel fill value after clipping. Metadata are handled according to observing-system type: satellite input metadata are transformed to geometry features and missing entries are filled using the available column mean in transformed space, while conventional metadata are filled using available column means and then mapped with fixed deterministic physical scalings. Target observations with missing required metadata are excluded from the target set rather than imputed. Radiosonde and aircraft pressure-level conditioning uses the nearest standard pressure level from the 16 configured pressure levels.

## 4. OCELOT Architecture

OCELOT follows a generalized encoder–processor–decoder (EPD) architecture (Figure 1) operating entirely in observation space and built around a shared latent spherical icosahedral mesh. As shown in Figure 1, heterogeneous observations from instrument groups (panel a) are first mapped onto the latent mesh by an instrument-specific bipartite GATv2 encoder (panel b; mechanism in panel g), then evolved by a hybrid processor that interleaves a sliding-window temporal Transformer with a spatial GNN mixer over the mesh edges (panel d; temporal mechanism in panel h). The evolved latent state is unrolled over four 3-hour steps and decoded back to each instrument's native observation space by a metadata-conditioned bipartite GATv2 decoder (panel e), producing forecasts at the original observation locations and lead times of +3, +6, +9, and +12 h (panel f). Key architectural hyperparameters are summarized in Table 2.

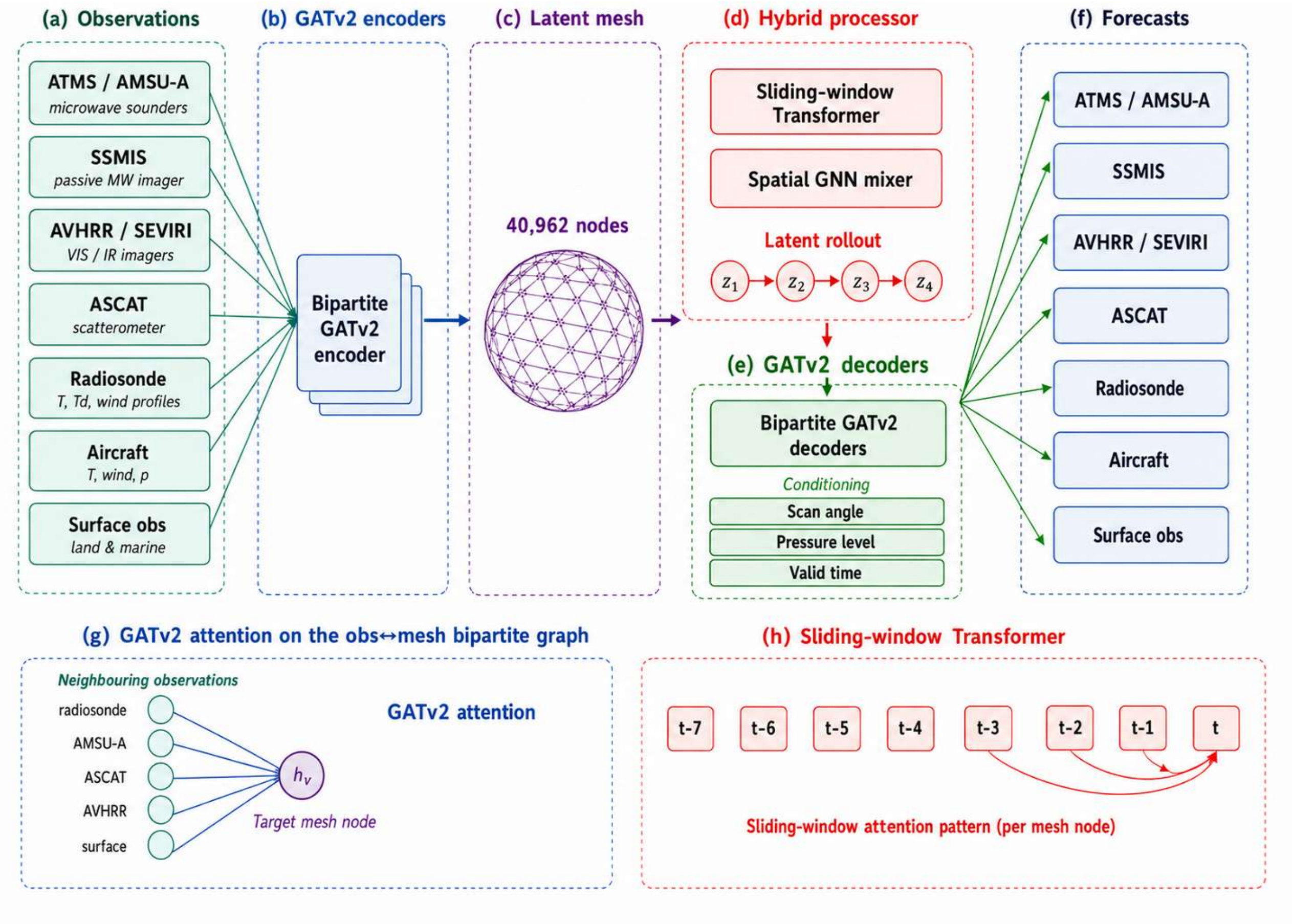

**Figure 1.** Schematic of the OCELOT v1 architecture for direct, multi-instrument observation forecasting. (a) Heterogeneous observations from microwave sounders, passive microwave imagers, VIS/IR imagers, ASCAT, radiosondes, aircraft reports, and surface networks enter the model in their native sensor geometry. (b)

Instrument-specific bipartite GATv2 encoders map observations onto a shared latent mesh; the attention mechanism is shown in (g). (c) The latent state is represented on a fixed icosahedral mesh with 40,962 nodes (k=6); multiscale edges are merged into a single graph. (d) A hybrid processor evolves the latent mesh through a sliding-window Transformer and spatial graph mixer over four 3-h latent rollout steps. (e) Instrument-specific GATv2 decoders map the latent mesh back to target observation locations using metadata-conditioned receiver queries. (f) Forecasts are produced at native observation locations for +3, +6, +9, and +12 h lead times. (g) GATv2 attention over the bipartite observation–mesh graph, using edge features based on relative position and edge length. (h) Sliding-window temporal attention with window length W=4, applied independently at each mesh node.

| **Component** | **Setting** |
|---|---|
| Mesh | Fixed icosahedral mesh; k = 6 refinements, 40,962 nodes, merged multiscale edges |
| Latent hidden dimension | 192 |
| Encoder | Bipartite GATv2, 2 layers, 4 heads, dropout 0.1, 4-D edge features |
| Decoder | Bipartite GATv2, 2 layers, 4 heads, dropout 0.1; metadata-conditioned receiver queries |
| Decoder metadata embeddings | Scan angle: 1→8→192; ASCAT scan angle: 3→8→192; pressure level: 16-entry embedding →8→192; target valid time: 5→8→192 |
| Processor | Hybrid sliding-window temporal Transformer + distance-weighted spatial graph mixer; depth 4, temporal window W = 4, 4 attention heads, hidden width 192, dropout 0.1 |
| Rollout | K=4 latent steps of $\Delta t$=3 h, decoded supervision at each step |
| Loss | Masked, instrument-rebalanced, channel-weighted MSE |
| Optimizer | Adam, learning rate = $1.5 \times 10^{-4}$, weight decay = $10^{-4}$, gradient clipping = 0.5 |
| LR schedule | Cosine annealing with 5% linear warmup, minimum learning rate = $10^{-6}$ |
| Mixed precision | 16-bit mixed precision with FP16 autocast and activation checkpointing |
| Distributed training | 4 nodes × 2 NVIDIA H100 GPUs; DDP with NCCL backend |

**Table 2.** Key OCELOT v1 model, optimization, and distributed-training settings

## 4.1 Per-Instrument Encoders and Observation Aggregation

OCELOT represents each observing system as its own set of observation nodes, allowing sensor-specific sampling density, geometry, metadata, and measured variables to be handled before fusion on the shared mesh. Because instruments differ widely in spatial density, sampling geometry, metadata structure, and physical variables, OCELOT assigns each instrument a dedicated encoder. Encoders transform raw measurements and metadata into latent embeddings using standardized normalization, geometric encodings, temporal features, and instrument identifiers.

Encoded observations are aggregated onto the latent mesh through a bipartite GATv2 layer (Brody et al. 2022). For each mesh node v, attention weights are computed over nearby observation nodes using 4-dimensional spatial edge features following the GraphCast convention: normalized great-circle distance and a relative-position vector (Lam et al. 2023). The attention mechanism is asymmetric and learnable, allowing the model to differentially weight observations within an instrument's footprint by metadata such as scan angle, polarization, or vertical position. Two stacked GATv2 message-passing rounds with 4 heads (output projected back to 192 dimensions) constitute the encoder.

## 4.2 Latent Mesh Representation and Geometry

All encoded observations are projected onto a shared unstructured spherical mesh that serves as OCELOT's latent state representation. In this study, OCELOT uses a fixed-resolution global icosahedral mesh constructed from (k=6) recursive triangle subdivisions of the icosahedron, yielding 40,962 nodes, with multiscale edges from intermediate refinements merged into a single graph following the GraphCast convention (Lam et al. 2023). The (k=6) resolution was selected as an engineering compromise between spatial expressiveness and the computational cost of global observation-to-mesh and mesh-to-observation message passing. It provides globally distributed latent support for the heterogeneous observation densities evaluated here while remaining tractable for repeated 3-h rollout, bipartite GATv2 encoding and decoding, and distributed mixed-precision training on the available H100 system.

Coarser meshes would reduce memory and communication cost but could over-smooth regional gradients and sparse-observation structure. Substantially finer meshes would increase edge counts, decoding cost, and graph-construction overhead, and their benefit has not yet been assessed through controlled mesh-resolution ablations in this first system paper. The fixed spherical mesh preserves global spatial coverage and provides a common latent domain for cross-instrument fusion. Mesh connectivity is fixed across forecast steps: the same mesh nodes and edges are used throughout the +3, +6, +9, and +12 h latent rollout, while the latent node states evolve in time. Each edge is associated with geometric features, including normalized distance and relative Cartesian direction, so that spatial message passing is informed by both separation and direction on the sphere. These edge features provide spatial context that helps the model learn spatial correlation and transport-like patterns without requiring the mesh topology to change during the forecast.

Expanding the latent graph into a three-dimensional mesh could explicitly represent vertical adjacency and may benefit tasks with dense, consistently leveled atmospheric profiles. We do not adopt that representation here because OCELOT inputs have heterogeneous vertical semantics, including surface variables, pressure levels, radiance channels, and column-integrated quantities. Instead, vertical position, channel identity, or instrument-specific vertical sensitivity is encoded through observation features and metadata during projection to the horizontal spherical mesh. This keeps the latent graph compact while allowing attention-based projection and decoding to learn instrument-specific vertical relationships.

### 4.3 Hybrid Sliding-Window Transformer–Spatial GNN Processor

The latent mesh state is evolved in time by a hybrid sliding-window Transformer / spatial-mixing processor (Figure 1d). The processor stacks four repeated blocks, each pairing temporal attention over recent latent states with spatial message passing on the icosahedral mesh.

First, a temporal multi-head self-attention block operates independently at each mesh node over a rolling cache of latent states. In this study, the cache length is W=4, matching the four 3-h latent rollout steps in the 12-h forecast horizon. With four attention heads and hidden width 192, the temporal block uses masked scaled dot-product attention,

$$Attn(Q, K, V) = softmax(QK^T / \sqrt{d_k} + M)V \quad (3)$$

Where Q, K, and V are the query, key, and value projections of the cached latent states at a given mesh node, $d_k$ is the key dimension, and M is the attention mask restricting attention to valid positions within the rolling window.

Second, a spatial graph-mixing block performs edge-weighted message passing along the merged multiscale icosahedral mesh edges. Spatial aggregation weights are computed from the same edge-feature convention used in the encoder and decoder: relative Cartesian position and edge length. The aggregated messages are passed through a learned MLP, followed by a residual connection and layer normalization.

This hybrid design combines temporal attention across recent latent states with mesh-local spatial mixing. The temporal block provides short-range memory across the rollout, while the spatial block imposes a geometric locality bias on information exchange over the spherical mesh. Hidden width is 192 throughout, with dropout 0.1, residual connections, and layer normalization. Activation checkpointing is applied to processor blocks to reduce peak memory during distributed mixed-precision training.

### 4.4 Per-Instrument Decoders with Query Conditioning by Projection

Instrument-specific decoders map latent mesh features back to predicted observations at native locations and times using bipartite GATv2 message passing over mesh-to-target graphs. Each decoder uses two GATv2 layers with four attention heads and dropout 0.1. Decoder weights are shared across rollout steps, so the same instrument-specific decoder is used for the +3, +6, +9, and +12 h forecasts.

For each target observation, the decoder receiver-node representation is initialized from projected metadata embeddings, a design we refer to as decoder-query conditioning by projection. The conditioning variables are embedded and projected to the full 192-dimensional latent space:

- Scan angle: scalar scan angle → 8-dimensional MLP embedding → 192-dimensional projection.
- Pressure level: 16-entry embedding lookup → 8-dimensional embedding →192-dimensional projection, used for radiosonde and aircraft targets.

- Target valid time: five calendar/time features →8-dimensional MLP embedding → 192-dimensional projection.
- ASCAT beam geometry: three scan-angle/beam components →8-dimensional MLP embedding → 192-dimensional projection.

Because all conditioning vectors are represented in the full 192-dimensional hidden space, decoder target nodes are initialized with metadata-aware queries before mesh-to-target message passing. This allows the decoder to condition predictions on viewing geometry, vertical coordinate, and valid time while using a shared latent mesh state.

## 5. Training Methodology

### 5.1 Data Sampling and Distributed Optimization

Training uses a fixed 12-hour input window and 12-hour target window, each comprising four 3-hour latent rollout steps. Each training step constructs only one bipartite graph for the current time bin, avoiding the pre-loading of long time sequences. In the current implementation, OCELOT is conditioned only on observations within the 12-hour input window; observations before that window are not carried forward through a recurrent state or longer history buffer.

We train OCELOT with PyTorch Lightning using Distributed Data Parallel across four nodes, each equipped with two NVIDIA H100 GPUs (8 GPUs total, NCCL backend). Each GPU processes a distinct temporal window, and gradients are all-reduced before each parameter update. Mixed-precision (16-mixed) arithmetic is used throughout, with gradient clipping at 0.5 to stabilize the attention-based processor. Activation (gradient) checkpointing is enabled to reduce peak GPU memory.

We optimize OCELOT with the Adam optimizer using an initial learning rate of $1.5 \times 10^{-4}$, weight decay of $10^{-4}$, and the dropout settings reported in Table 2. The learning rate follows a cosine-annealing schedule with linear warmup over the first 5% of training steps, starting from a warmup factor of $10^{-2}$ and decaying to a minimum learning rate of $10^{-6}$.

### 5.2 Multi-Step Latent Rollout and Supervision

For each time bin, OCELOT produces forecasts at +3, +6, +9, and +12 h through latent rollout. At each rollout step, the latent mesh state is decoded to the target variables and supervised against the verifying target at that lead time; the total loss is averaged over the four lead times. Predicted values are used only for loss computation and are not fed back into the encoder, which keeps the training objective well posed and avoids scheduled-sampling complications.

5.3 Loss Function

OCELOT is trained with a masked, instrument-rebalanced, channel-weighted mean-squared-error objective applied directly in observation space. For rollout step k, instrument i, channel c, and observation j, the per-element squared error in standardized units is:

$$e_{kicj} = (\hat{y}_{kicj} - y_{kicj})^2 \tag{4}$$

A binary validity mask ($m_{kicj} \in \{0,1\}$) excludes missing values, quality-control-failed observations, and inactive target elements. Let $w_{ic} \geq 0$ denote a configurable per-instrument, per-channel weight. Channels with zero weight are inactive. The per-instrument weighted loss at rollout step $k$ is:

$$L_i^k = \Sigma_{c,j}\, w_{ic}\, m_{kicj}\, e_{kicj} \,/\, (\Sigma_{c,j}\, m_{kicj}\, 1(w_{ic} > 0) + \varepsilon) \tag{5}$$

where ε is a small numerical floor. Thus, channel weights scale the squared-error contribution in the numerator, while the denominator counts valid, nonzero-weight target elements. Because observing systems differ by orders of magnitude in sample count, we apply instrument rebalancing. Let $I_k$ denote the set of instruments with at least one valid target element at rollout step $k$. The step-wise loss is:

$$L^k = (1 \,/\, |I_k|)\, \Sigma_{i \in Ik}\, L_i^k \tag{6}$$

The total multi-step rollout loss is computed as a lead-time-weighted average:

$$L = \Sigma_{k=1}{}^K\, \lambda_k\, L^k \,/\, \Sigma_{k=1}{}^K\, \lambda_k \tag{7}$$

Here $K$=4 and $\lambda_k$ is the lead-time weight. We set all $\lambda_k$=1, so the objective reduces to an average over the four 3-h rollout steps. In implementation, instrument-step terms with no valid target elements are skipped. Setting all $w_{ic}$=1 and disabling instrument rebalancing recovers the standard masked global MSE over all valid target elements. Nonuniform channel weights allow selected channels or variables to be emphasized or de-emphasized while preserving masking and instrument-level rebalancing. A Huber variant is implemented but is not used in this study; in our training runs, the weighted MSE objective was more stable in mixed precision and produced lower validation error at this model size.

### 5.4 Model-Development Variants

During model development, we evaluated multiple loss, encoder/decoder, and processor variants before selecting the OCELOT v1 configuration reported here. These experiments informed the choice of the masked, instrument-rebalanced, channel-weighted MSE objective over an earlier weighted Huber formulation. The Huber variant improved robustness in some settings but tended to downweight physically meaningful large wind errors as outliers, whereas the weighted MSE objective produced more stable mixed-precision training and stronger validation performance in our development runs.

We also evaluated several modular encoder, decoder, and processor variants. Interaction-network-based encoder/decoder designs were tested before adopting bipartite GATv2 layers, which better supported adaptive weighting over nearby observations and mesh nodes. For the processor, early interaction-network variants were computationally more expensive; sliding-window Transformer variants improved temporal handling; and the final OCELOT v1 processor combines sliding-window temporal attention with mesh-local spatial mixing. These alternatives remain implemented in the OCELOT codebase. This paper evaluates the final OCELOT v1 system; controlled ablations quantifying each modular design choice are left for future work.

### 5.5 Computational Implementation and Cost Reporting

The production training configuration uses eight NVIDIA H100 GPUs across four nodes, with Distributed Data Parallel, mixed precision, activation checkpointing, windowed Zarr reads, and per-bin graph construction. These implementation choices are necessary because the cost is dominated not only by neural-network FLOPs, but also by observation-archive I/O, graph construction, and mesh-to-target decoding over heterogeneous target sets. Table 2 reports the hardware, optimizer, precision, rollout, and distributed-training settings required to reproduce the configuration.

We do not report a definitive training-time, throughput, or cost-per-forecast benchmark for OCELOT v1 because wall-clock performance depends strongly on the storage backend, cache state, subsampling settings, validation-output frequency, and active target families. We therefore treat computational-cost characterization as incomplete in this first system paper rather than inferring cost from a single development run. A standardized scalability benchmark, including training wall time, samples per second, graph-construction throughput, inference latency, and memory scaling with mesh resolution and observation count, is identified as a priority for future releases.

## 6. Evaluation Framework and Results

We evaluate OCELOT v1 using the held-out 2025 period, with two forecast initializations per day, at 00 UTC and 12 UTC, for 730 total initializations across the year. This annual evaluation spans all four seasons and is used for both native observation-space verification and diagnostic comparisons with the operational NCEP Global Forecast System (GFS).

The primary evaluation is performed in native observation space. For each observing system, OCELOT decoders produce forecasts at the original target observation locations, valid times, and metadata coordinates. Metrics are then computed directly against the corresponding held-out measurements after inverse transformation to physical units. This preserves the native sampling of each observing system and avoids gridding or interpolation of the verification observations.

We also perform mesh-space diagnostics by querying the same decoders at the fixed icosahedral mesh nodes. This produces globally distributed OCELOT fields with one value per mesh node and lead time. For observation-location forecast comparisons, GFS forecast fields at the same initialization time and forecast hour are interpolated to the matched observation locations. For mesh-space diagnostics, GFS analysis fields valid at the forecast verification time are interpolated to the OCELOT mesh nodes. These two diagnostics serve different purposes: the first provides an operational forecast reference, while the second evaluates the spatial realism of OCELOT's mesh-query output against an analysis-based gridded reference.

Unless otherwise stated, RMSE is computed as a count-weighted metric over all matched samples, and confidence intervals are estimated by bootstrap resampling over forecast initializations with replacement. For the surface persistence baseline, we use the nearest prior same-location observation of the same variable within the 12-hour input window. To keep the comparison fair, OCELOT, GFS, and persistence are evaluated on the common subset of target observations for which this persistence match is available.

For temperature variables, all error metrics reported in the text are expressed in K because RMSE and bias are temperature differences; plotted temperature fields may still be shown in °C where noted. Wind-component errors are reported in m $s^{-1}$.

## 6.1 Observation-Space Forecast Skill at 12 Hours

Across instrument families, OCELOT achieves stable, physically plausible 12-hour forecast skill when evaluated over the full 2025 forecast set. Passive microwave temperature-sounding channels show the lowest errors and consistent behavior across independent instruments, while the infrared AVHRR window channel is more challenging because of its stronger sensitivity to surface temperature, clouds, and sub-footprint variability. Conventional temperature profiles also show coherent vertical structure, with radiosonde and aircraft forecasts closely following the observed mean profiles and exhibiting smooth pressure-dependent RMSE.

Near-surface variables remain stable through the 12-hour rollout. Surface 2-m air-temperature RMSE increases only modestly from about 3.2 K at +3 h to about 3.6 K at +12 h. This is higher than GFS, as expected for an operational cycled data-assimilation system, but substantially lower than persistence at longer leads, which grows to more than 6 K by +12 h. The same pattern holds for 10-m wind components: OCELOT errors increase gradually with lead time and remain below persistence at +12 h, although GFS retains the lowest RMSE. Because OCELOT produces the +3, +6, +9, and +12 h forecasts from a shared latent rollout with independent supervised heads, the +12 h skill reflects a direct supervised forecast rather than autoregressive feedback of intermediate predictions.

## 6.2 Spatial Structure of Forecast Skill in Observation Space

Figure 2 examines whether +12 h errors remain geographically coherent across independent satellite observing systems. Panels (a)–(c) show three passive microwave temperature-sounding channels in the 50–55 GHz oxygen absorption band: ATMS channel 7, AMSU-A channel 7, and SSMIS channel 5. Despite differences in platform, scan geometry, and sampling pattern, the three channels show broadly consistent large-scale brightness-temperature structure and spatially coherent residuals. Their pooled +12 h RMSE values are 1.87 K for ATMS, 1.24 K for AMSU-A, and 1.54 K for SSMIS, supporting the interpretation that OCELOT is learning shared atmospheric-layer information rather than instrument-specific sampling artifacts.

Panel (d) shows the AVHRR infrared window channel, which is more sensitive to surface temperature, cloud-top emission, and sub-footprint variability. This channel has a larger pooled +12 h RMSE of 3.95 K, but still captures the dominant spatial structures, including warm continental regions, marine cloud regimes, and cold high-latitude areas. The larger and more localized residuals are consistent with the greater physical heterogeneity of infrared window-channel observations. The swath-aligned structures visible in the panels reflect the native orbital sampling within the observation window and are not, by themselves, indicative of model error.

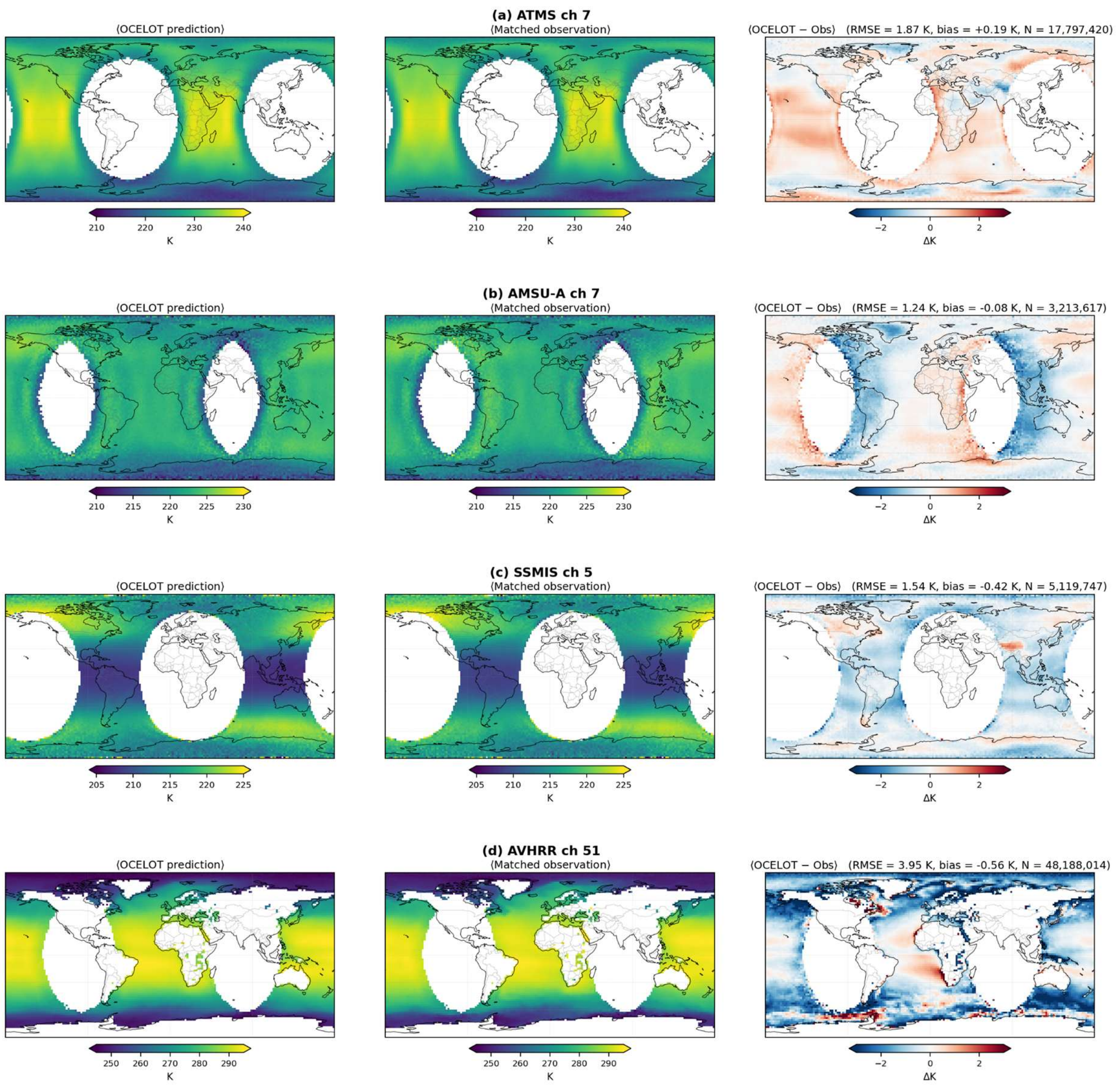

**Figure 2.** Multi-instrument observation-space error structure of OCELOT +12 h forecasts, pooled across all 2025 forecast initializations from the 00 UTC and 12 UTC cycles and verified at +12 h. For each instrument, panels show the mean OCELOT prediction (left), mean matched observation (centre), and mean residual (OCELOT − observation; right), aggregated onto a 2° × 2° latitude–longitude grid across all matched samples. Residual panels share a common symmetric ±3 K colour scale. Per-row RMSE and bias values are computed from the unbinned matched prediction–observation pairs before spatial aggregation. Rows: (a) ATMS ch 7, (b) AMSU-A ch 7, (c) SSMIS ch 5, (d) AVHRR infrared window channel.

### 6.3 Vertical Profile Forecasts for Aircraft and Radiosondes

Figure 3 evaluates the vertical structure of OCELOT observation-space temperature forecasts using all 2025 radiosonde and aircraft prediction files from the 00 UTC and 12 UTC initialization cycles. Profiles are aggregated over the 0-12 h forecast window using the nominal 3, 6, 9, and 12 h lead times. The top row compares the pooled mean matched observations with the corresponding OCELOT predictions, while the bottom row shows pooled RMSE by pressure level. Grey horizontal bars indicate the number of matched observations contributing to each level.

For radiosonde temperature, OCELOT closely follows the observed vertical temperature structure from the surface to 10 hPa. RMSE ranges from about 3.2 to 5.2 K, with the lowest errors in the upper stratospheric levels around 30-100 hPa and larger errors near 10 hPa and in parts of the troposphere. The bias remains modest, ranging from approximately -1.3 to +1.0 K. OCELOT shows a small warm bias in the lower troposphere, transitions to a weak cold bias through much of the mid- and upper troposphere/lower stratosphere, and returns to a small warm bias at the highest levels. The smooth vertical variation of both the mean profile and RMSE suggests that the pressure-level conditioning captures coherent vertical dependencies in the observation-space decoder.

Aircraft temperature profiles show similarly consistent structure over the reported sampling range from 1000 to 200 hPa. Pooled RMSE ranges from about 3.5 to 4.8 K, with the lowest errors in the lower troposphere and the largest values near 500 hPa. The aircraft profile has a modest warm bias below roughly 700 hPa, near-neutral bias around 500-300 hPa, and a weak cold bias near 250-200 hPa. The aircraft column is restricted to pressures at or below 200 hPa because sampling above this level is too sparse for stable level-wise statistics. The much larger aircraft sample sizes at common flight levels help produce a relatively smooth RMSE profile across the troposphere.

Across both radiosonde and aircraft temperature profiles, RMSE varies smoothly with pressure, with no abrupt layer-wise discontinuities. This indicates that OCELOT learns vertically coherent relationships and maintains stable coupling across pressure levels within the observation-space forecasting framework.

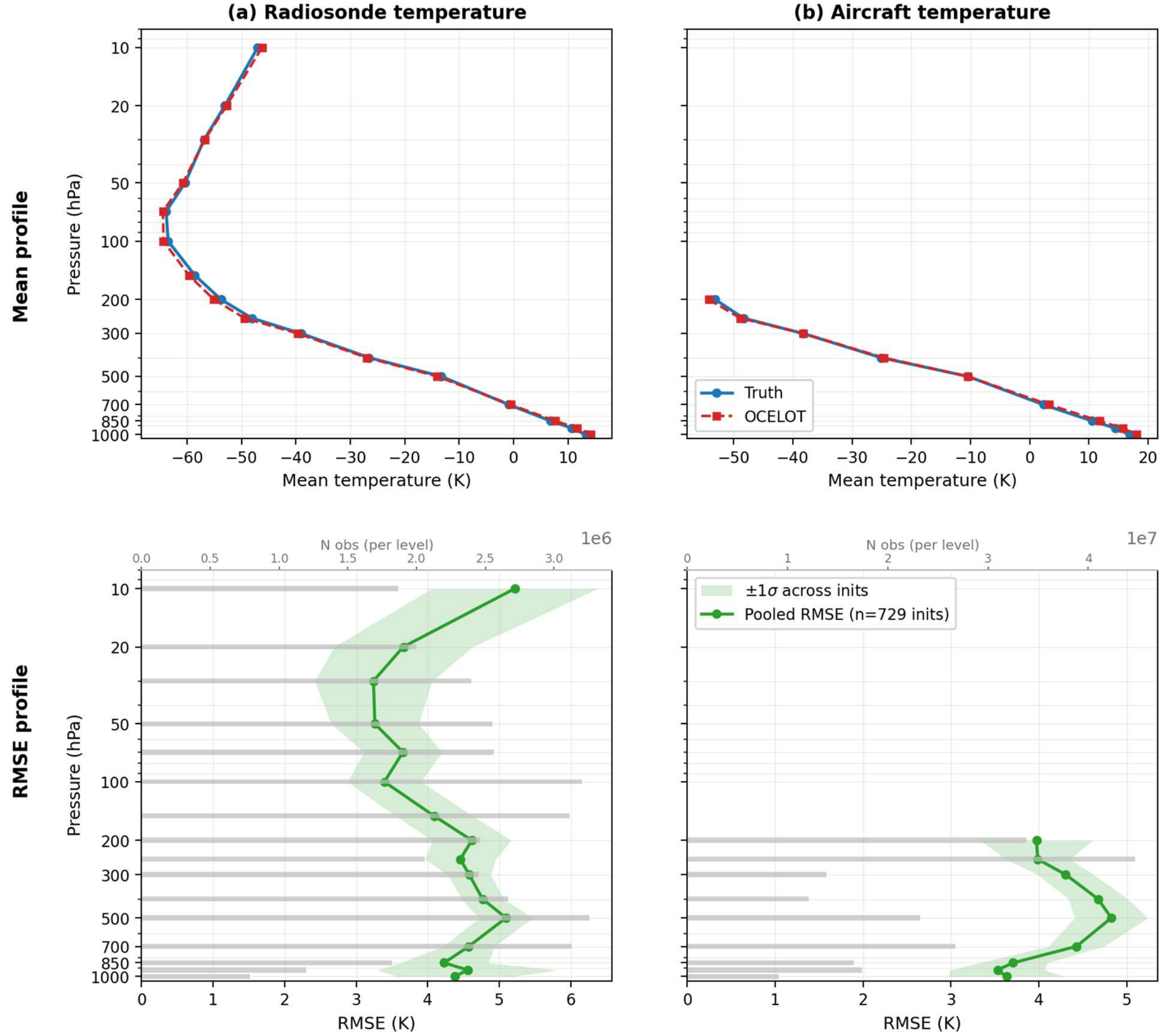

**Figure 3.** Vertical profile diagnostics for OCELOT observation-space forecasts, pooled across the full held-out 2025 evaluation set, consisting of 730 forecast initializations from the 00 UTC and 12 UTC cycles, and aggregated over the 0–12 h forecast window. Top row: mean profile of matched observations (solid blue) versus OCELOT predictions (dashed red). Bottom row: pooled RMSE (solid green) with ±1σ across-initialization shading; grey bars on the upper axis indicate the number of matched observations per pressure level. Columns: (a) radiosonde temperature, (b) aircraft temperature.

## 6.4 Temporal Rollout Diagnostics

To quantify temporal error growth, Figure 4a–d show the spatial structure of OCELOT minus observation differences in surface air temperature at +3, +6, +9, and +12 h for the 2025-04-25 00 UTC initialization, using a common symmetric ±10 K color scale. Figure 4e shows the corresponding lead-time evolution of error pooled across the 2025 00 UTC and 12 UTC initialization cycles.

The spatial patterns indicate that forecast error growth occurs primarily through gradual amplification of coherent regional structures rather than the emergence of new large-scale biases. Residuals remain geographically organized throughout the rollout, with localized warm and cold structures over land regions and generally weaker, more diffuse residuals over oceanic areas. The error field remains physically interpretable from +3 to +12 h.

For the map initialization, RMSE increases from 2.74 K at +3 h to 3.23 K at +12 h, with intermediate values of 2.76 K and 2.96 K at +6 h and +9 h, respectively. The corresponding bias remains positive but modest, decreasing from +0.58 K at +3 h to +0.19 K at +9 h before increasing to +0.45 K at +12 h.

Because all lead-time predictions are produced from a single shared latent rollout with independent supervised heads, this behavior reflects increasing forecast difficulty with lead time rather than instability from autoregressive feedback. Within the evaluated 12-h horizon, we find no evidence of rapid spatial-noise amplification or short-range rollout instability. The combination of smooth RMSE growth, small bias, and persistent spatial coherence indicates that the latent-space evolution remains stable and well conditioned over the 12-hour horizon.

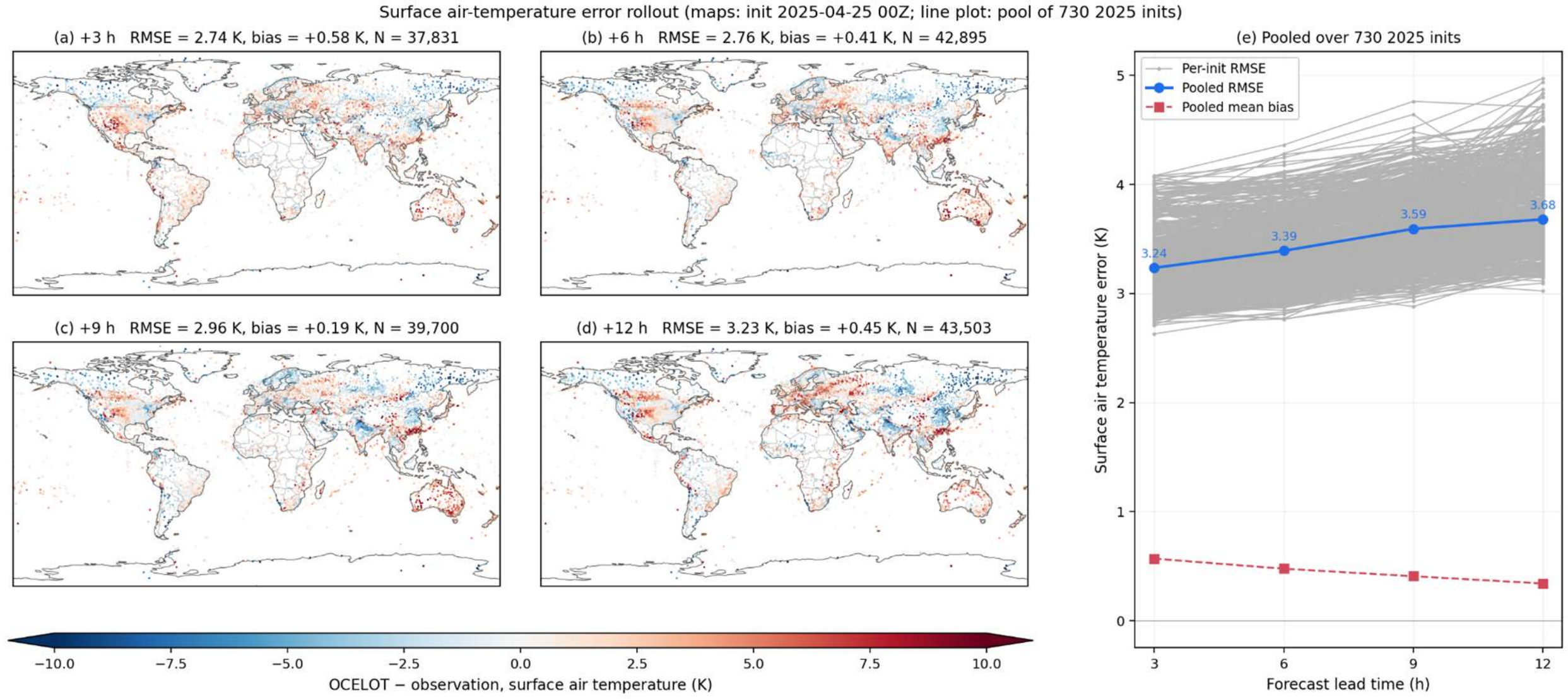

**Figure 4.** Temporal rollout of surface 2-m air-temperature forecast errors. Panels (a)-(d) show OCELOT minus observation differences at +3, +6, +9, and +12 h for initialization 2025-04-25 00 UTC, using a common symmetric +/-10 K color scale. Panel (e) shows the lead-time evolution of error across the 2025 00 UTC and 12 UTC initialization cycles: thin grey lines denote per-initialization RMSE, the solid blue curve shows count-weighted pooled RMSE, and the dashed red curve shows pooled mean bias.

### 6.5 Diagnostic Comparison Against the Operational GFS

To place OCELOT's surface forecasts in an operational context, we compare observation-location predictions against the NCEP Global Forecast System (GFS, 0.25°) forecast fields interpolated to the matched observation locations. GFS is used here as a mature operational physics-based NWP reference initialized through cycled data assimilation, not as a like-for-like machine-learning baseline. We separately use GFS analysis fields for mesh-space diagnostics in Figure 7 to assess the spatial realism of OCELOT's mesh-query output at the valid time. Because GFS benefits from cycled data assimilation and the analysis assimilates observations near the verification time, these comparisons should not be interpreted as operational parity.

Figure 5 presents observation-space verification for surface 2-m air temperature at +12 h for both 00 UTC and 12 UTC initializations on 2025-04-23. OCELOT reproduces the large-scale observed temperature distribution in both cycles, including the dominant warm continental regions, cold high-latitude areas, and broad land-ocean gradients. The residual panels show that OCELOT errors are spatially coherent rather than noise-like, with warm residuals over parts of Eurasia and regional mixed-sign errors over North America and the tropics. GFS residuals are generally smaller in magnitude, as expected for an operational assimilative forecast system.

For the 00 UTC initialization, OCELOT has RMSE = 3.22 °C and bias = +0.47 °C, compared with GFS RMSE = 2.57 °C and bias = −0.41 °C. For the 12 UTC initialization, OCELOT has RMSE = 3.28 °C and bias = +0.40 °C, compared with GFS RMSE = 2.50 °C and bias = −0.72 °C. Thus, in these two representative +12 h cases, OCELOT trails GFS by about 0.7–0.8 °C in RMSE while maintaining comparable large-scale spatial structure.

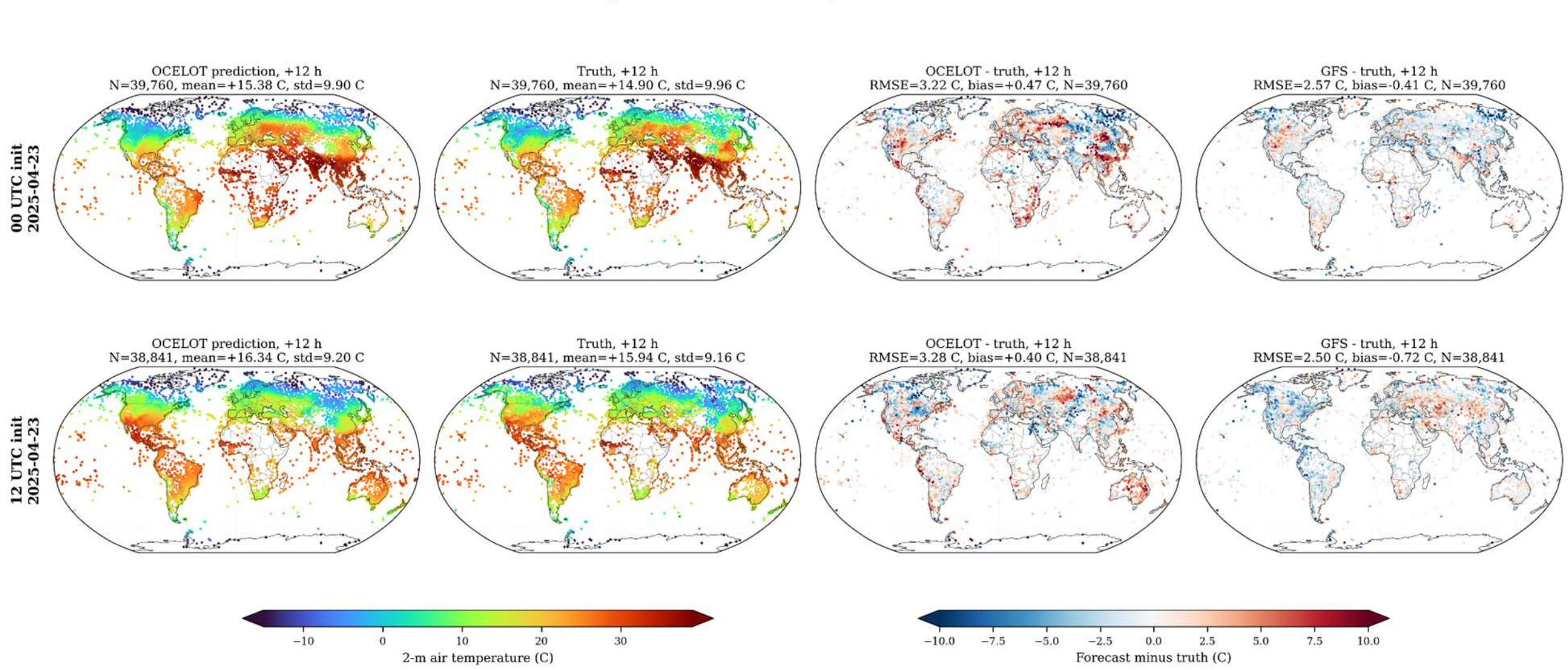

**Figure 5.** Observation-space comparison of surface 2-m air temperature at +12 h for the 2025-04-23 00 UTC and 12 UTC forecast initializations. Rows show the 00 UTC initialization verified at 2025-04-23 12 UTC and the 12 UTC initialization verified at 2025-04-24 00 UTC. Columns show OCELOT prediction, matched surface-observation truth, OCELOT − truth residual, and GFS − truth residual, with GFS interpolated to the matched observation locations. Field panels share a common temperature color scale, and residual panels share a common symmetric error color scale. Field-panel statistics report sample size, mean, and standard deviation; residual-panel statistics report RMSE, bias, and sample size computed from the matched observation-space pairs.

Across all 2025 GFS-comparison initializations, Figure 6 compares OCELOT, GFS, and a same-location persistence baseline for surface 2-m air temperature and 10-m wind components on the common persistence-valid subset. The persistence reference is intentionally simple: it uses the nearest prior same-location observation and therefore provides a no-learning lower-bound benchmark rather than a strong statistical forecasting system. For 2-m air temperature, OCELOT RMSE increases from 3.23 K at +3 h to 3.64 K at +12 h, while GFS remains lower, increasing slightly from 2.75 to 2.83 K. The persistence baseline is competitive at +3 h (2.51 K) but degrades rapidly with lead time, reaching 6.66 K by +12 h, consistent with the lack of diurnal-cycle correction in a simple same-location persistence forecast. Stronger observation-only baselines, including climatology and autoregressive station-level models, are useful future additions but are not included in this first v1 evaluation.

For 10-m winds, OCELOT is close to GFS at short lead times and remains substantially better than persistence at longer lead times. At +12 h, OCELOT RMSE is 2.53 m s−1 for zonal wind and 2.61 m s−1 for meridional wind, compared with 2.19 and 2.16 m s−1 for GFS and 3.23 and 3.15 m s−1 for persistence, respectively. These results show that OCELOT provides meaningful short-range skill beyond a no-learning persistence reference, while GFS retains lower RMSE for the surface variables at later lead times.

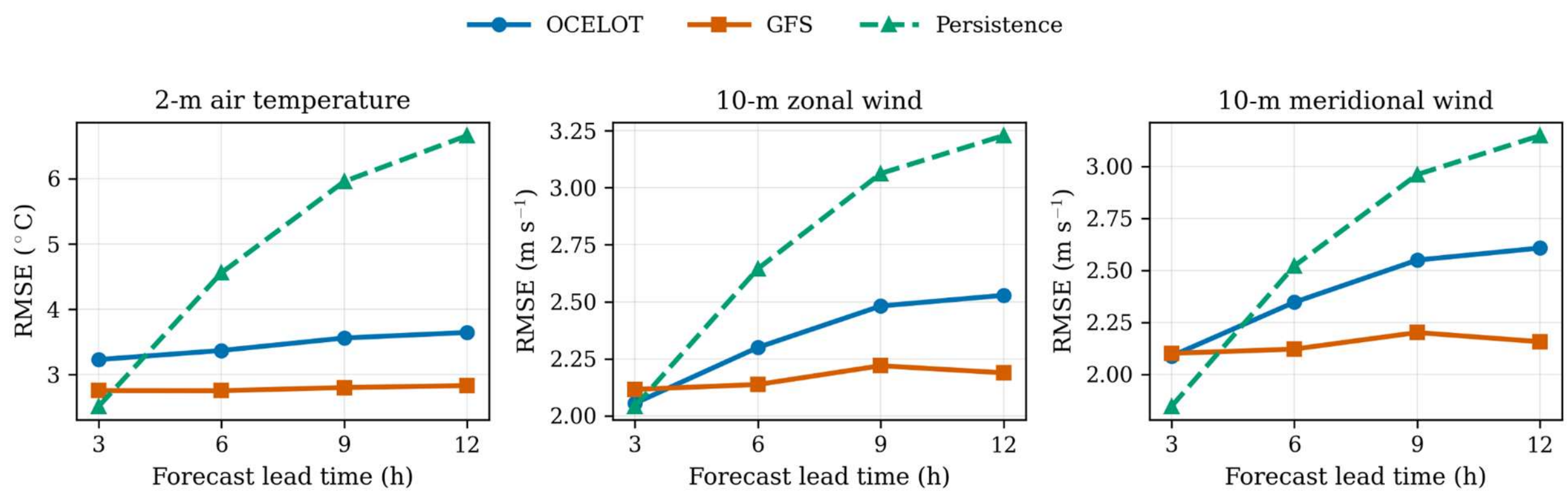

**Figure 6.** Surface-observation RMSE for OCELOT, GFS, and a same-location persistence baseline, pooled over 730 2025 GFS-comparison initializations. Panels show 2-m air temperature, 10-m zonal wind, and 10-m meridional wind at +3, +6, +9, and +12 h lead times. All methods are evaluated on the same subset of observations for which the persistence baseline is available. Persistence uses the nearest prior same-location observation and does not correct for the diurnal cycle, which is especially relevant for 2-m temperature.

Figure 7 provides a gridded-analysis diagnostic of OCELOT's mesh-query output using the same-date 00 UTC and 12 UTC +12 h forecasts. OCELOT is decoded on the fixed icosahedral mesh and compared with the GFS analysis interpolated to the same mesh nodes at the valid time. In both cycles, OCELOT captures the dominant large-scale 2-m temperature structure, including the meridional temperature gradient, warm tropical and subtropical regions, continental thermal structure, and cold high-latitude areas.

The OCELOT-minus-analysis residuals are spatially organized rather than noise-like. The 00 UTC and 12 UTC examples have mesh-space RMSE values of 3.37 K and 3.21 K, respectively, with positive biases of +0.76 K and +0.57 K. Larger differences occur regionally, especially over complex land areas, high-latitude regions, and parts of the Southern Ocean, while broad hemispheric-scale temperature gradients remain well represented. This diagnostic indicates that the observation-trained latent mesh supports physically coherent gridded outputs, even though OCELOT is trained and verified primarily in native observation space. Because the GFS analysis assimilates observations near the valid time, this comparison is used only as a gridded-analysis diagnostic, not as a forecast-vs-forecast benchmark.

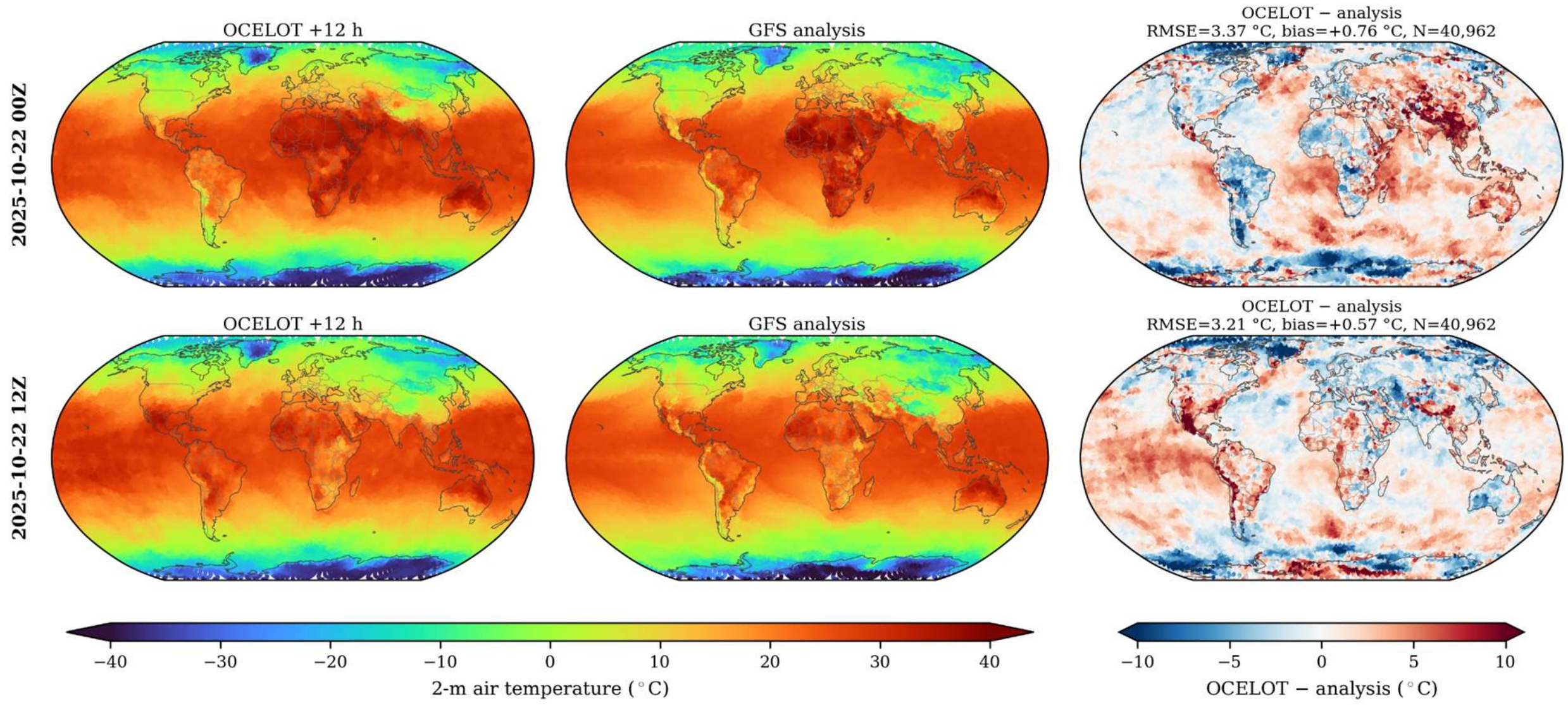

**Figure 7.** Example same-date 00 UTC and 12 UTC mesh-space diagnostics for OCELOT 2-m air-temperature forecasts verified against the GFS analysis. OCELOT is decoded on the fixed icosahedral mesh at +12 h, and the GFS analysis valid at the forecast verification time is interpolated to the same mesh nodes. Columns show OCELOT, GFS analysis at valid time, and OCELOT − analysis. Rows show the 2025-10-22 00 UTC and 12 UTC forecast initializations. Residual-panel statistics are computed over all 40,962 mesh nodes.

Together, these diagnostics show that OCELOT recovers physically coherent large-scale surface-temperature structure, improves over persistence at longer lead times, and produces spatially organized mesh-query fields, while GFS retains lower RMSE for surface variables.

## 7. Discussion

This study demonstrates the feasibility of short-range global atmospheric forecasting directly in observation space across heterogeneous satellite and in-situ observing systems. OCELOT learns a coherent representation of atmospheric evolution without relying on gridded model states, reanalysis targets, or variational data assimilation, instead operating on native observations and their associated metadata.

A central result is that the shared latent icosahedral mesh supports effective fusion of heterogeneous observations while preserving instrument-specific geometry. Across the full 2025 evaluation, OCELOT produces spatially coherent forecasts for microwave sounders, infrared imagers, conventional profiles, and surface variables. The consistency among independent microwave temperature-sounding instruments, together with the smooth radiosonde and aircraft profile errors, suggests that the model captures shared atmospheric structure rather than fitting each observing system in isolation.

Temporal rollout diagnostics further show that the learned dynamics remain stable over the 12-hour forecast horizon. For surface 2-m air temperature, pooled RMSE increases gradually from about 3.2 K at +3 h to about 3.6–3.7 K at +12 h, with no evidence of rapid spatial-noise amplification. Forecast errors evolve primarily through gradual amplification of coherent regional structures rather than the emergence of unstable small-scale noise, suggesting that the latent-space evolution is well conditioned over the evaluated horizon.

The comparison against GFS and persistence provides an operational reference for OCELOT's feasibility as a standalone observation-space system. OCELOT currently yields higher RMSE than the operational GFS for surface variables, which is a reflection of the GFS's mature cycled data-assimilation system, explicit land-surface modeling, and extensive physical-model development. However, OCELOT substantially outperforms persistence at longer lead times for 2-m temperature and 10-m wind components. This suggests that the learned latent evolution captures meaningful short-range atmospheric change beyond a static initial state.

The surface-variable gap likely stems from OCELOT v1 lacking static mesh descriptors (elevation, land-sea masks) and certain surface-sensitive microwave channels. These missing constraints force the model to infer terrain and boundary-layer effects indirectly, contributing to larger regional errors compared to GFS.

The hybrid sliding-window Transformer/spatial-GNN processor was designed to combine short-range temporal memory with mesh-local spatial information exchange. Temporal attention provides a mechanism for coupling the four 3-hour latent rollout steps, while the spatial graph component introduces a geometric locality bias over the spherical mesh. The stable 12-hour rollouts reported here show that this combined design can be trained and evaluated without evident short-horizon instability. In preliminary processor comparisons, the hybrid processor outperformed a sliding-window-only configuration, suggesting that explicit mesh-local spatial mixing improves forecast skill. This development path suggests that explicit mesh-local spatial mixing is beneficial for OCELOT v1, while a fully controlled ablation study of individual architectural components remains outside the scope of this first system paper.

Decoder conditioning by projection is a critical design choice for handling heterogeneous target metadata. By projecting embeddings for scan angle, pressure level, target valid time, and instrument-specific geometry into the full hidden dimension, the decoder initializes target-node queries in a metadata-aware latent space before mesh-to-target message passing. This conditioning is particularly important for vertical profile observations, where pressure level defines the target's vertical context, and for cross-track scanning radiometers, where scan position modulates the observed brightness temperature through viewing-geometry effects. During model development, simpler decoder initializations that did not include this projected metadata conditioning produced physically inconsistent satellite and radiosonde predictions, indicating that target metadata must be represented explicitly at decode time. The results presented here therefore support decoder-query conditioning as an essential component of OCELOT v1.

Beyond performance, the observation-centric formulation offers several conceptual advantages. By training and verifying directly against native measurements, OCELOT avoids requiring a gridded analysis target and preserves instrument-specific geometry throughout the learning problem. This allows cross-instrument relationships to be learned without first projecting all observations onto a fixed state vector. The framework also provides a natural route for adding new observing systems through instrument-specific encoders and decoders while maintaining a shared latent atmospheric representation. Mesh-query outputs provide an additional bridge to traditional NWP diagnostics and workflows, while the primary learning objective remains observation based.

## 8. Limitations and Future Directions

### 8.1 Current Limitations

The current study is limited to a 12-hour forecast horizon and latent rollouts consisting of four 3-hour steps. While sufficient to test the fusion of heterogeneous data into a stable forecast, it does not yet establish medium-range predictive skill or fully separate learned dynamics from short-term memory. Extending OCELOT beyond 12 hours will require additional temporal modeling and evaluation against stronger persistence, climatology, and autoregressive baselines.

Direct comparison with other operational and machine-learning systems remains an ongoing effort. While comparisons with systems like ECMWF IFS, GraphCast, or Pangu-Weather are essential for context, creating the necessary observation-matched test sets for these grid-trained models is a significant engineering and data-alignment challenge. A controlled head-to-head comparison with GraphDOP is also a priority, requiring aligned archives, target variables, and verification protocols.

Physical consistency and statistical characterization also require further study. Current diagnostics suggest physically coherent large-scale structure but do not yet rigorously test conservation laws, geostrophic balance, or moisture budgets. Additionally, while bootstrap resampling provides initial uncertainty estimates, more comprehensive evaluations are needed to account for spatial dependencies, seasonal regime changes, and calibrated probabilistic performance.

Finally, the computational-cost discussion remains incomplete. While OCELOT v1 uses distributed mixed-precision training and efficient I/O, a systematic scalability study is still required to quantify training wall time, inference latency, and memory scaling as a function of mesh resolution and observation density.

### 8.2 Planned Future Directions

Future work will focus on improving both the scalability and predictive capability of the OCELOT framework. Because the largest remaining performance gap occurs for near-surface variables, near-term development will target improved representation of land-surface and boundary-layer processes. The results reported here do not yet include static mesh features such as elevation or land–sea mask, nor do they include additional surface-sensitive satellite observations such as AMSR passive microwave imager channels. These are being incorporated in ongoing development. Elevation and land–sea mask will provide explicit information about terrain, coastlines, and land–ocean contrast, which are central to 2-m temperature and 10-m wind errors. Surface-sensitive microwave channels can provide additional constraints on lower-atmospheric and surface conditions, helping the latent state better represent boundary-layer and land-surface variability. Together, these additions are intended to reduce regional surface-temperature and near-surface wind errors, particularly over complex terrain, coastal zones, and regions with strong diurnal or boundary-layer variability.

Beyond these surface-focused improvements, future work will also address model scalability, longer forecast horizons, uncertainty quantification, and high-impact-event behavior. One direction is to activate and benchmark the implemented latent-hierarchical mesh formulation (40,962 → 10,242 → 2,562 → 642 nodes), including U-Net–style cross-scale conditioning, to enable scale-aware information propagation. Another priority is extending the rollout horizon beyond 12 hours and introducing probabilistic outputs with calibrated uncertainty estimates. Probabilistic or extreme-aware objectives may also help reduce smoothing of rare, high-gradient features, particularly when small displacement errors would be strongly penalized by deterministic MSE. Targeted evaluation on tropical cyclones, atmospheric rivers, and extreme temperature or precipitation episodes will be important for assessing model behavior in dynamically complex regimes.

A further direction is to use the trained model for Forecast Sensitivity-based Observation Impact (FSOI), building on the observation-space FSOI infrastructure included in the OCELOT codebase. Because OCELOT operates directly on observations, it provides a natural setting for diagnosing observation impact without first mapping observations into a gridded analysis state. Finally, systematic evaluation of mesh- and grid-projected OCELOT products against multiple NWP and ML baselines will be important for operational interoperability, while preserving the observation-centric training formulation.

## 9. Conclusions

We have presented OCELOT, an observation-centric framework for short-range global atmospheric forecasting that operates directly on heterogeneous satellite and in-situ measurements. Building on the emerging observation-space forecasting paradigm, OCELOT learns the temporal evolution of observations in their native sensor geometry through a shared latent icosahedral mesh, per-instrument bipartite GATv2 encoders and decoders, a hybrid sliding-window Transformer/spatial-GNN processor, and decoder-query conditioning by projection of metadata into the full hidden dimension.

Out-of-sample evaluation on held-out 2025 observations demonstrates that short-range forecasting in observation space is feasible and stable at global scale. Across the full 2025 evaluation set, consisting of 730 forecast initializations from the 00 UTC and 12 UTC cycles, OCELOT produces coherent forecasts across diverse observing systems, including passive microwave sounders, infrared imagers, radiosondes, aircraft, and surface observations. At +12 h, selected microwave temperature-sounding channels achieve RMSE values of 1.24–1.87 K, while the more surface- and cloud-sensitive AVHRR infrared window channel shows a higher RMSE of 3.95 K. Conventional temperature profiles show vertically coherent errors, and surface 2-m air-temperature forecasts remain stable through the 12 h horizon, with pooled RMSE increasing from about 3.2 K at +3 h to about 3.6 K at +12 h.

Diagnostic comparison with GFS and persistence establishes OCELOT's utility within an operational context. While GFS retains lower RMSE for surface variables by leveraging cycled data assimilation and mature physical parameterizations, OCELOT demonstrates significant skill by substantially outperforming a same-location persistence baseline at longer lead times. This result confirms that the observation-centric framework captures physically coherent atmospheric evolution. Mesh-space diagnostics further illustrate that OCELOT produces spatially organized structures that align with operational analysis, supporting its potential as a reanalysis-independent forecasting pathway.

OCELOT demonstrates that observation-space models can recover physically coherent atmospheric structures without reanalysis targets. Future work will address current gaps through static mesh features, surface-sensitive observations, and extended probabilistic rollouts for high-impact events.

**Acknowledgments**

This work was supported by the National Oceanic and Atmospheric Administration (NOAA). Computational resources were provided by [NOAA HPC system, Ursa]. The development of OCELOT was inspired in part by the GraphDOP team's pioneering work on observation-space forecasting.

**Data Availability Statement**

The OCELOT model source code, configuration files, and scripts used for training, inference, and evaluation are publicly available from the NOAA-EMC/ocelot GitHub repository at the fixed release tag ocelot-v1.0 (https://github.com/NOAA-EMC/ocelot/releases/tag/ocelot-v1.0) and are distributed under the GNU Lesser General Public License v2.1 (LGPL-2.1). The preprocessed observation archives, GFS fields used for diagnostic comparison, trained model weights, and evaluation outputs supporting the figures are available from the corresponding author upon reasonable request.